\documentclass[aps,pre,longbibliography,twocolumn,floatfix]{revtex4-1}
\usepackage{graphicx}
\usepackage{amsmath}
\usepackage{amssymb}
\usepackage{bm}
\usepackage[pdftex]{hyperref}
\hypersetup{colorlinks=true,linkcolor=blue,citecolor=blue,urlcolor=blue}
\usepackage{verbatim}

\newcommand{\expect}[1]{\langle {#1}\rangle}

\begin{document}

\title{The Gardner transition in physical dimensions}
\author{C.~L.~Hicks}
\author{M.~J.~Wheatley}
\author{M.~J.~Godfrey}
\author{M.~A.~Moore}
\affiliation{School of Physics and Astronomy, University of Manchester,
Manchester M13 9PL, UK}

\date{\today}

\begin{abstract}
The Gardner transition is the transition that at mean-field level
separates a stable glass phase from a marginally stable phase.  This
transition has similarities with the de Almeida-Thouless transition of
spin glasses.  We have studied a well-understood problem, that of
disks moving in a narrow channel, which shows many features usually
associated with the Gardner transition.  We show that some of these
features are artifacts that arise when a disk escapes its local cage
during the quench to higher densities.  There is evidence that the
Gardner transition becomes an avoided transition, in that the
correlation length becomes quite large, of order 15 particle
diameters, even in our quasi-one-dimensional system.
\end{abstract}
\maketitle

In a remarkable series of papers (see Ref.~\cite{Charbonneau:16a} for
a review and references) the large-dimension limit of the hard sphere
fluid has been solved.  This program of calculation provides the
mean-field description relevant for the dynamical glass phase
transition, the Kauzmann ideal glass transition, the Gardner
transition and the geometrical description of the properties of jammed
states.  The next step is of course to understand what happens in
finite dimensions.  In this paper we argue that at least the Gardner
transition is not a real transition in physical dimensions, $d\le3$.
The Gardner transition is the transition associated with the emergence
of a complex free-energy landscape composed of many marginally stable
sub-basins contained within a larger glass metabasin \cite{seoane:18}.  It is thus
similar to a state of a spin glass in a phase with broken replica
symmetry \cite{Urbani:15}.  In fact the field theory of the Gardner
transition is closely related to that of the Ising spin glass problem
in a random magnetic field -- the de Almeida-Thouless transition
\cite{Charbonneau:15}.

It has been argued for some time that the de Almeida-Thouless
transition only occurs in systems with dimensions $d > 6$
\cite{moore:11, wang:17, moore:17}, but this is still controversial \cite{mattsson:95,joerg:08a,baity:14,baityjanus:14}.
Furthermore in two dimensions there is not even a spin glass
transition in zero field, and none has been detected either in a
finite field.  Nevertheless, both simulations \cite{Berthier:16} and
experiments on hard disks in two dimensions \cite{Seguin:16,Ohern:16}
seem to provide evidence for a Gardner transition during compression;
i.e., that glass systems that start with the same particle positions
but with different particle velocities, can, by the end of the
compression, be in mutually inaccessible states.  To understand what
might be going on, we have applied the methods of
Ref.~\cite{Berthier:16} to a system of hard disks moving in a narrow
channel, which is a model that has previously been studied in some
detail
\cite{Robinson:16,Godfrey:15,Godfrey:14,bowles:06,Yamchi:12,Ashwin:13,
  Yamchi:15,Ashwin:09, Kofke:93, Varga:11, Gurin:13}.  Our system,
being one dimensional, cannot have a true phase transition.  We find
that we can produce the same kinds of behavior that other
investigators studying two- and three-dimensional hard sphere systems
have explained in terms of the state-following Gardner transition.
However, because of the simplicity of our system we can give a full
explanation of our observations.  We find that some of the behavior
that has been ascribed to the Gardner transition (such as a nontrivial
change in the distribution of an order parameter) arises when the time
scales associated with the quench used in the state-following
investigations are long enough for there to be a significant chance of
a disk escaping its cage by crossing the channel.  Our results thus do
not imply that there are glasses that are stable and glasses that are
marginal, with the Gardner transition separating the two
\cite{Berthier:16}.  Effects that have been attributed to the Gardner
transition in two- and three-dimensional hard-sphere systems may also
have a simple explanation, connected with just a few disks or spheres
escaping their cages on the time scale of the quench.  In a recent
preprint, Scalliet et al.~\cite{scalliet:17} have invoked a similar
mechanism to help explain the absence of a Gardner transition in a
system of particles that interact via a soft potential.  However, for
our system of hard disks we can provide evidence that there is an
avoided Gardner transition at which the correlation length grows to a
large value, ${\approx}15$ particle diameters, but does not diverge.
For hard spheres in three dimensions, which is closer to the
mean-field limit of infinite dimension, we suspect that the equivalent
length scale will be larger still and exceed the linear dimensions of
the systems that were simulated \cite{Berthier:16}, making it
impossible to distinguish a true phase transition from an avoided one.

\begin{figure}
  \includegraphics[width=\columnwidth]{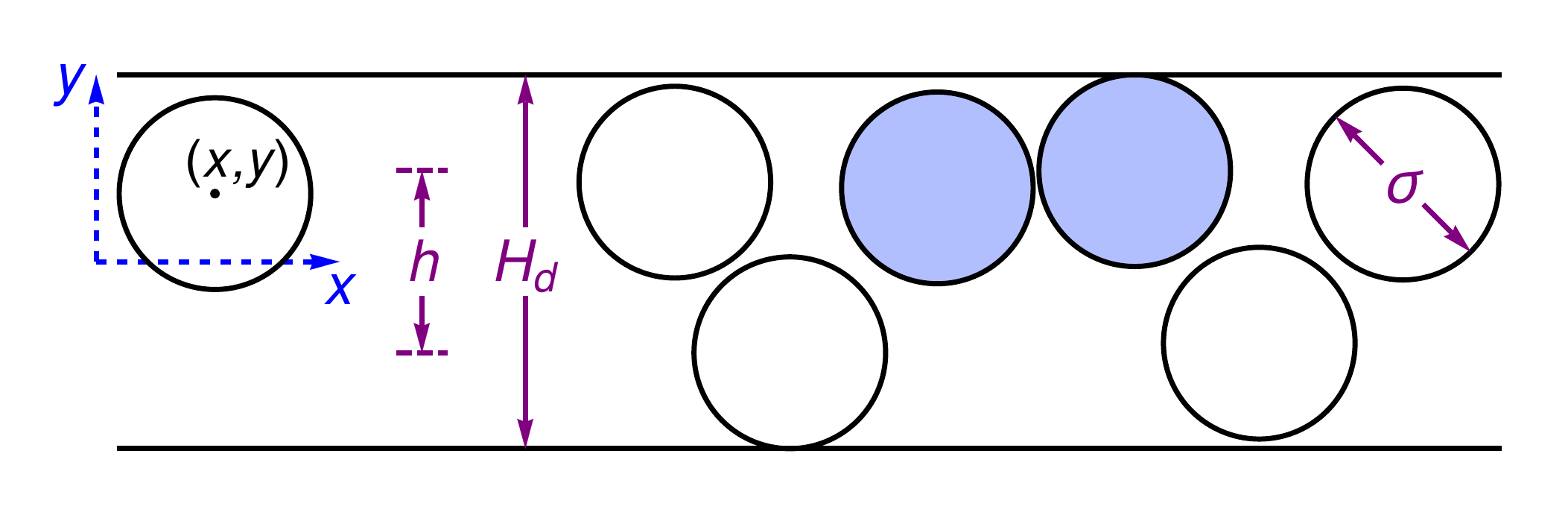}
  \caption{(Color online) The system of hard disks in a channel.  The
    distance $H_d$ is the width of the channel, $\sigma$ is the
    diameter of each disk, and $h=H_d-\sigma$ is the width of the
    channel accessible to the centers of the disks.  For the
    coordinates $(x,y)$ of the disk, $y$ is measured from the center
    line of the channel.  The blue shaded disks can be regarded as a
    \textit{defect} in the zigzag arrangement of the disks that is
    favored at high density.}
  \label{fig:notation}
\end{figure}

Details of the system that we are studying are given in
Fig.~\ref{fig:notation}.  The packing fraction $\phi$ is defined as
$\phi=N\pi\sigma^2/(4 H_d L)$, where $N$ is the number of disks in a
channel of length $L$.  The channel width $H_d$ was taken to be
$1.95\sigma$, where $\sigma$ is the diameter of a disk; we have
previously made extensive transfer-matrix \cite{Baxter:82}
calculations of thermodynamic properties and correlation functions for
this case~\cite{Godfrey:15}.

The dynamics in this system start to slow as ``zigzag'' order sets in
above a packing fraction $\phi =\phi_d\approx 0.48$
\cite{Godfrey:14,Godfrey:15,Robinson:16}.  This kind of order is
characterized by successive values of $y_i$ taking opposite signs (see
Fig.~\ref{fig:notation}).  The zigzag order can be interrupted by
defects where successive $y_i$ are of the same sign; the correlation
length $\xi$ for zigzag order is approximately half the average
distance between defects~\cite{Godfrey:14}.  These defects play an
important role in our analysis of the dynamics of the system. (Defects
of one kind or another seem to play an important role across the whole
of glass physics \cite{wyart:17,keys:11,ritort:03,stillinger:88}.)
Their spacing increases rapidly with increasing $\phi$, such that
$\xi$ passes $2000$ at $\phi=0.7206$ and reaches $\xi=2.3\times 10^6$
at $\phi=0.76$.

We use event-driven molecular dynamics for our simulations.  The
number of disks $N$ is taken to be $4000$ and periodic boundary
conditions are applied in the $x$-direction.  We start the system in
an initial ``equilibrium'' state of packing fraction $\phi_i\ge 0.70$
and use the Lubachevsky-Stillinger algorithm \cite{LS:90} to compress
it to values of $\phi>\phi_i$ on a time scale much less than the
$\alpha$ relaxation time at the packing fraction $\phi_i$.  We put the
word ``equilibrium'' in quotes as for the packing fractions studied
there should be no or very few defects present in the system, but we
have found there are typically ${\sim}10$ present, owing to imperfect
equilibration.  During the compression, the diameter of the disks is
increased at a rate $\dot{\sigma}$ and the width of the channel is
also increased so that the relation $H_d=1.95\sigma$ is maintained
throughout the compression~\cite{Yamchi:12}.  Such a compression is
``state following'' if on the time scale of the quench most of the
disks remain caged and do not move far from their initial positions.
The kinetic energy of the disks increases during the compression, so,
after the quench, the disks are assigned new velocities drawn from a
Maxwellian distribution for particles of mass $m=1$ at a reciprocal
temperature $\beta=1$.  The unit of length is chosen so that
$\sigma=1$.

\begin{figure}
  \includegraphics[width=\columnwidth]{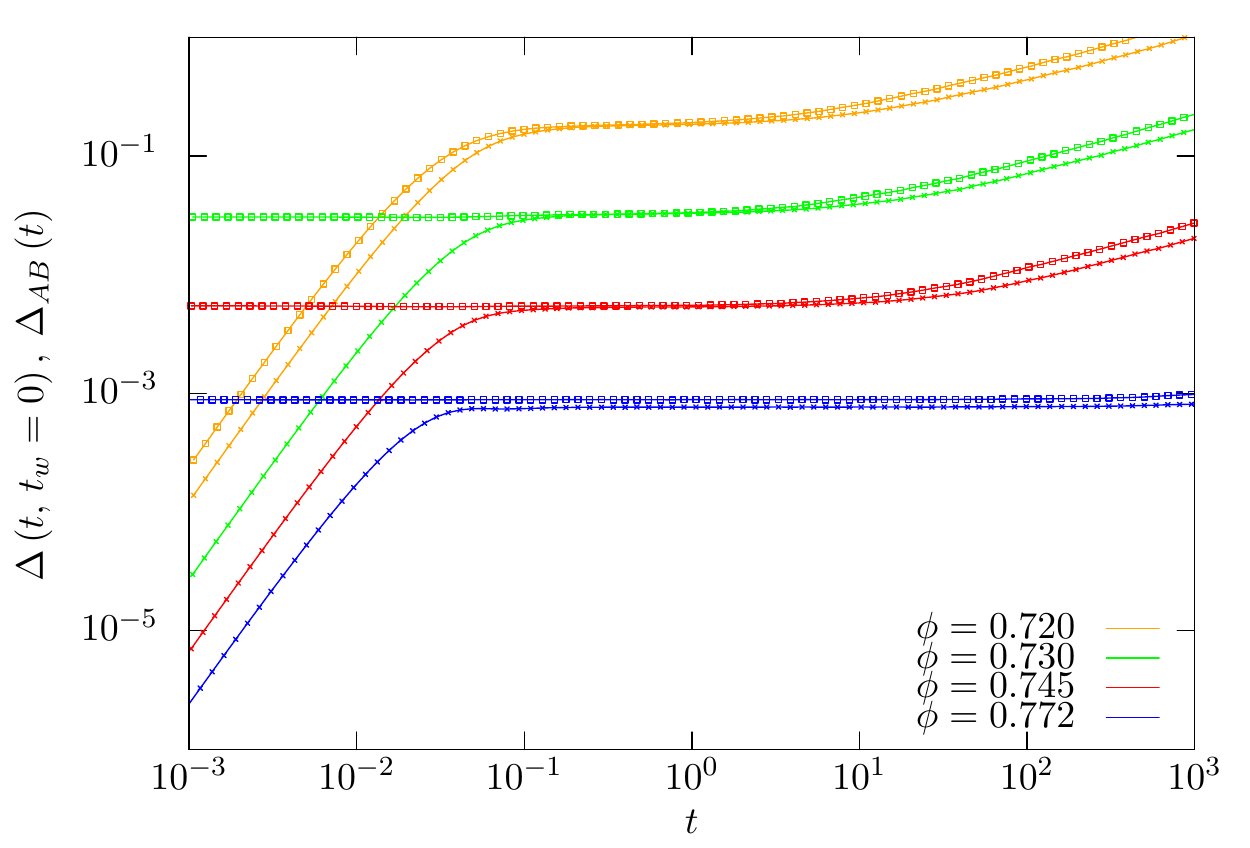}
  \caption{(Color online) Mean-squared displacements $\Delta(t,t_W=0)$
    [crosses] and $\Delta_{AB}(t)$ [squares] as a function of time
    $t$, where the quench was started from the initial ``equilibrium''
    state of packing fraction $\phi_i= 0.720$, which contains defects,
    and compressed to packing fractions $\phi=0.730$, $0.745$ and
    $0.772$.  Data have been multiplied by $5^k$, with $k=0$, \dots,
    $3$ for $\phi=0.772$, \dots, $0.72$ respectively so the data
    points do not obscure each other.  Note that for
    $\phi=\phi_i=0.720$ there was no quench; no disks crossed the
    channel in this case, so $\Delta_{AB}$ follows $\Delta$ closely.}
  \label{Deltas}
\end{figure}

\begin{figure}
  \includegraphics[width=\columnwidth]{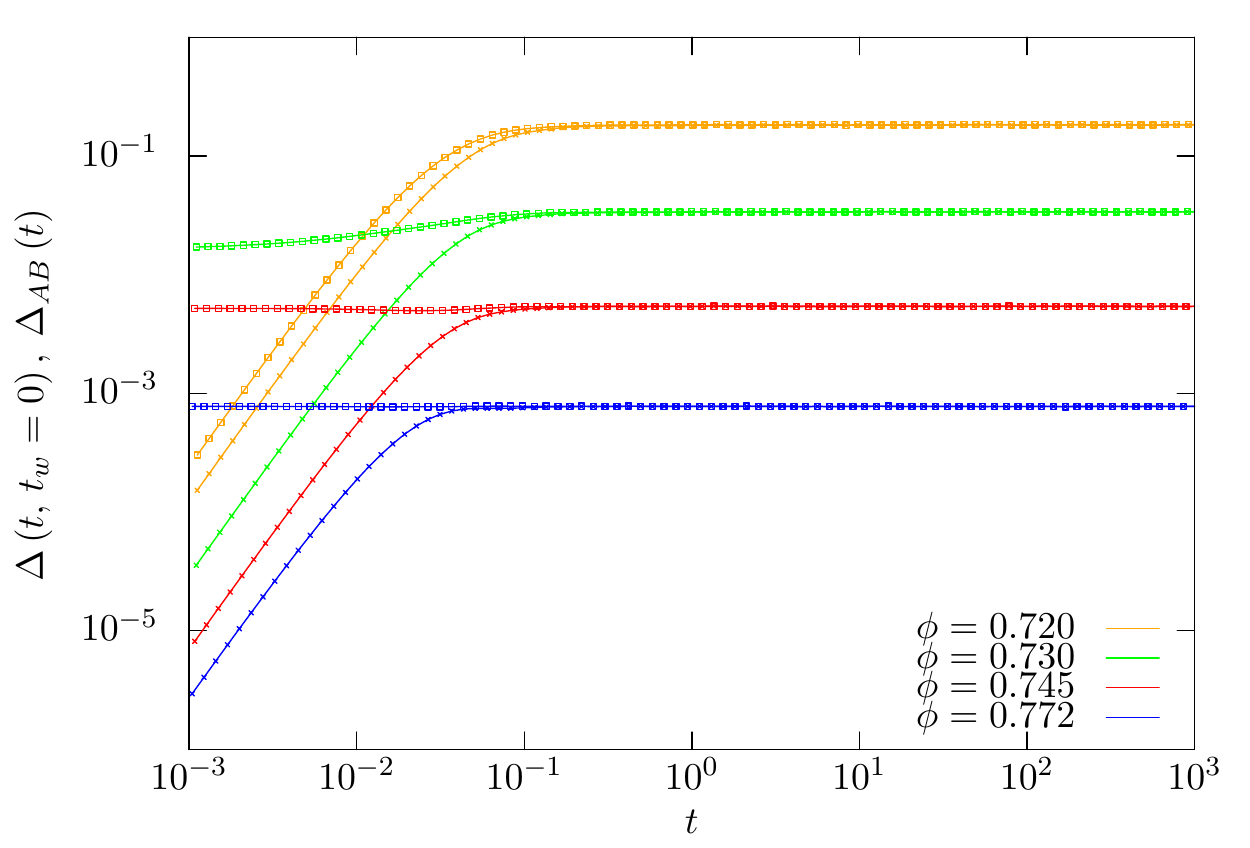}
  \caption{(Color online) Mean-squared displacements $\Delta(t,t_W=0)$
    [crosses] and $\Delta_{AB}(t)$ [squares] as a function of time
    $t$, where the quench was started from an initial state of packing
    fraction $\phi_i= 0.720$, which contained \emph{no defects}, and
    compressed to packing fractions $0.730,0.745$ and $0.772$.  Data
    have been multiplied by $5^k$, with $k=0$, \dots, $3$ for
    $\phi=0.772$, \dots, $0.72$, respectively, so the data points do
    not obscure each other.  Notice that there is no splitting on the
    plateau between $\Delta$ and $\Delta_{AB}$.}
  \label{Deltasnodefects}
\end{figure}

\begin{figure}
  \includegraphics[width=\columnwidth]{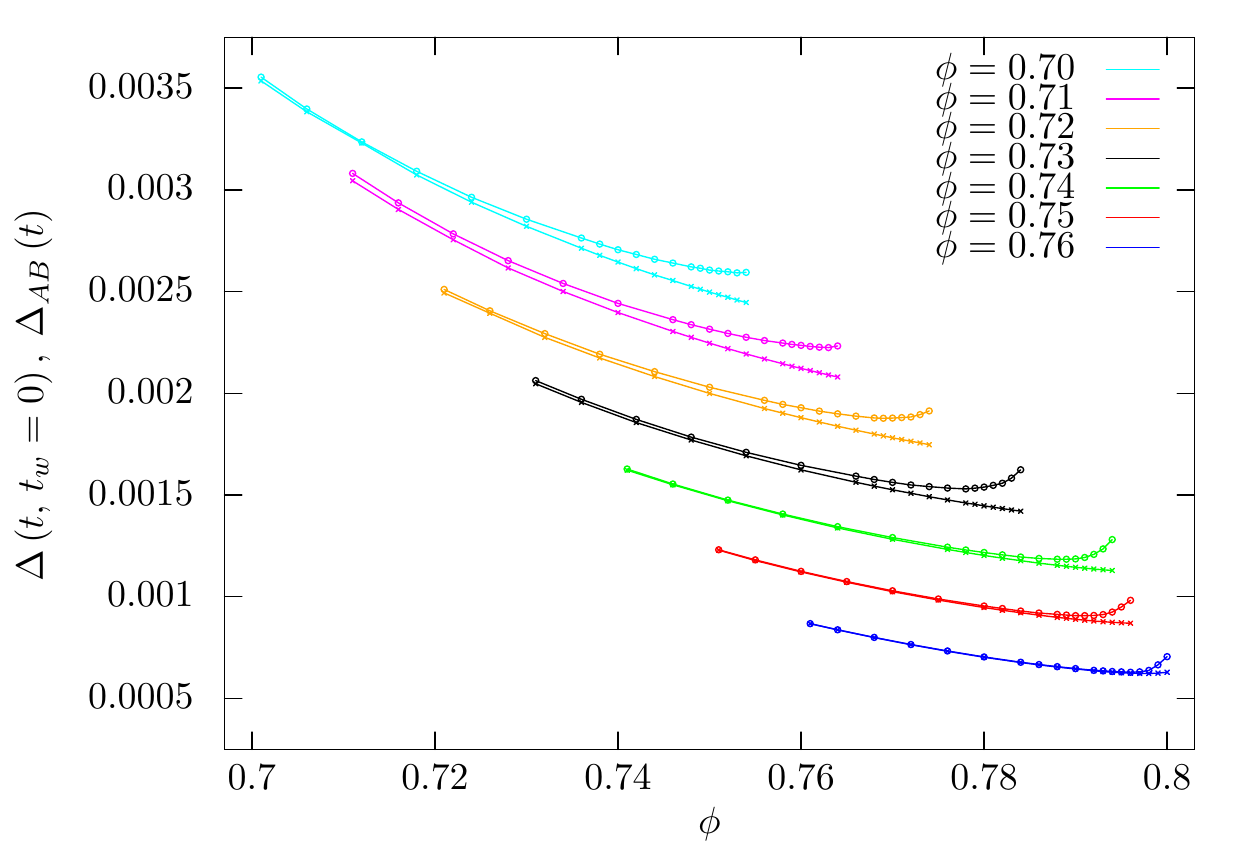}
  \caption{(Color online) Comparison of the plateau values of $\Delta$
    and $\Delta_{AB}$ for the system with defects showing that the two
    separate, with $\Delta_{AB}>\Delta$, as $\phi$ increases.  For
    clarity, the data have been shifted up by $0.00025 k$, where
    $k=0$, $1$, \dots, $6$ for $\phi=0.76$, $0.75$, \dots, $0.70$,
    respectively.  Error bars are insignificant in comparison to the
    size of the data points.}
  \label{Deltasvsphi}
\end{figure}

\begin{figure}
  \includegraphics[width=\columnwidth]{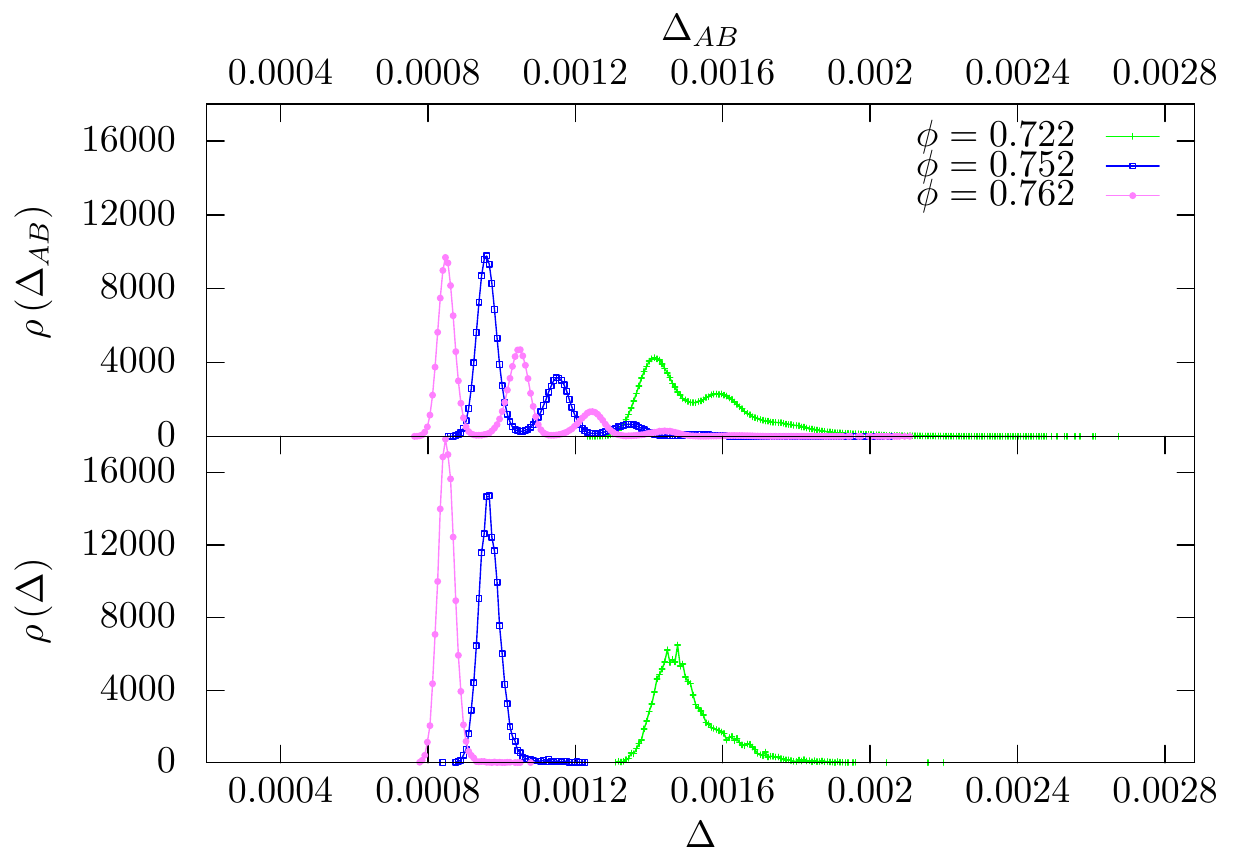}
  \caption{(Color online) Distribution of plateau values of $\Delta$,
    $\rho(\Delta)$ and of $\Delta_{AB}$, $\rho(\Delta_{AB})$ following
    a quench from $\phi_i=0.71$ to the packing fractions $\phi=0.722$,
    $0.752$ and $0.762$.  The smaller peaks in $\rho(\Delta_{AB})$ are
    due to one or more disks crossing the channel during the quench.}
  \label{Histo071}
\end{figure}

We judge the extent of caging by studying the mean-squared
$y$-displacements (MSD) of the disks from their initial positions
$y_i(t_W)$,
\begin{equation}
  \Delta(t,t_W)=\frac{1}{N}\sum_{i=1}^N\langle[y_i(t+t_W)-y_i(t_W)]^2\rangle,
  \label{Deltadef}
\end{equation}
averaged over both thermal fluctuations and initial states.  The time
$t$ starts after a waiting time $t_W$ when the compression has reached
a chosen target packing fraction $\phi$.  The MSD $\Delta(t,t_W)$
grows initially as $t^2$ from zero at $t=0$ (see Fig.~\ref{Deltas}).
For our choice of channel width the disks cannot pass each other so
the ordering $0 \le x_1 < x_2 < \cdots < x_N < L$ is preserved at all
times, and the disks are always caged in the $x$-direction.  Caging in
the $y$-direction is indicated by the presence of a plateau in
$\Delta(t, t_W=0)$; the plateau is present for packing fractions
$\phi> 0.7$ (see Fig.~\ref{Deltas}), rather than for
$\phi\gtrsim\phi_d$, as seen in two- and three-dimensional systems.
Caging is a pre-requisite for seeing the state-following Gardner
transition \cite{Berthier:16}.  In contrast to the studies of two- and
three-dimensional systems, $\Delta(t,t_W)$ cannot increase
indefinitely as $t\to\infty$ since in our system $\lvert y_i\rvert\le
h/2$.  The $\alpha$ relaxation time $\tau_\alpha$ is defined as the
time at which $\Delta$ leaves the plateau, which in our system is the
result of disks crossing from one side of the channel to the other;
quantitative estimates for $\tau_\alpha$ are given in the Supplemental
Material \cite{Suppl:TMMethods}.  Channel crossing is catalyzed by the
presence of the defects; if defects are absent, they must first be
nucleated in pairs \cite{Godfrey:15,Robinson:16}, and this is a
process that takes much longer than the $\alpha$ relaxation time
\cite{Robinson:16}.  We have prepared initial states free of defects
and have found for such states that we cannot observe the end of the
plateau on the times for which we could run the simulation; this is
shown in Fig.~\ref{Deltasnodefects}.  In such systems there is no
channel crossing on accessible time scales.

To see Gardner-like behavior, one must study independent copies A and
B of the system \cite{Berthier:16}.  Initially, the disks in A and B
have identical coordinates $(x_i, y_i)$, taken from the
``equilibrium'' state at $\phi_i$, but they are assigned different
velocities drawn from a Maxwellian distribution.  The quantity studied
is the MSD of their separation,
\begin{equation}
  \Delta_{AB}(t)=\frac{1}{N}\sum_{i=1}^N\langle[y_{A,i}(t)-y_{B,i}(t)]^2\rangle.
\label{DeltadefAB}
\end{equation}
In Fig.~\ref{Deltas} we show our results for $\Delta$ and
$\Delta_{AB}$ as a function of time for various initial packing
fractions.  At the larger packing fractions, they show a feature
that has been previously been regarded as a signature of the Gardner
transition, a difference between the plateau values of $\Delta(t,t_W)$
and $\Delta_{AB}(t)$ \cite{Berthier:16}.  These have been plotted
against $\phi$ in Fig.~\ref{Deltasvsphi}.  To extract the plateau
values, we used the method of Ref.~\cite{Berthier:16}, in which the
MSD starting from a given equilibrium state are rescaled to a
universal curve of $\Delta/\Delta_m$ against $t/t_m$, where $\Delta_m$
and $t_m$ are $\phi$-dependent scaling factors.

Plateau values of $\Delta(t,t_W)$ and $\Delta_{AB}(t)$ can also be
calculated on the assumption that no disk crosses the channel, so that
in Eq.~(\ref{Deltadef}), $y_i(t+t_W)$ and $y_i(t_W)$ take the same
sign; similarly, $y_{A,i}(t)$ and $y_{B,i}(t)$ take the same sign in
Eq.~(\ref{DeltadefAB}).  Then, on the plateau, $\Delta(t,t_W)$ and
$\Delta_{AB}(t)$ can be approximated by
\begin{equation}
  \Delta(t,t_W)= \Delta_{AB}(t) \approx 2(\langle y^2\rangle-\langle\lvert y\rvert\rangle^2).
  \label{Deltapredict}
\end{equation}
We calculate $\langle y^2\rangle$ and $\langle\lvert y\rvert
\rangle^2$ using the transfer-matrix method of Ref.~\cite{Godfrey:15};
the results from Eq.~(\ref{Deltapredict}) are indistinguishable from
the plateau values of $\Delta$ and $\Delta_{AB}$ seen in
Fig.~\ref{Deltasnodefects}.

In recent work, Scalliet et al. \cite{scalliet:17} have shown that a
splitting of the plateau values of $\Delta$ and $\Delta_{AB}$ is not
of itself sufficient to show that there is a state-following Gardner
transition that requires the stable glass to split into
marginally-stable sub-basins where the two copies A and B sometimes
end up.  The split can instead be due to a small subset of the
particles being frozen into slightly different locations in the two
copies.  As we shall see, an explanation of this kind also applies to
our system of disks in a channel.

For very large times, $t\gg\tau_\alpha$, disks may cross the channel,
so that $\langle y_i(t+t_W) y_i(t_W) \rangle\to 0$; thus,
$\Delta(t,t_W)$ and $\Delta_{AB}(t)$ will both reach a \emph{second}
plateau at the equilibrium value, $2\langle y_i^2\rangle\approx
h^2/2$.  However, this long-time limit is not relevant to the
state-following Garder transition, where one is interested in times
such that $\Delta(t,t_W)$ and $\Delta_{AB}(t)$ are still on their
first plateau.  At the $\alpha$ relaxation time disks begin to escape
their cages by crossing from one side of the channel to the other.
Channel crossing is strongly suppressed in the absence of defects;
thus, for a system with no defects there can be no splitting in the
first plateau values of $\Delta$ and $\Delta_{AB}$ at large packing
fractions; this is confirmed by the results shown in
Fig.~\ref{Deltasnodefects}.  When defects are present, $\Delta_{AB}$
and $\Delta$ separate, as shown in Fig.~\ref{Deltasvsphi}.


To understand the cause of the splitting we turn to the histograms of
the plateau values of $\Delta$ and $\Delta_{AB}$ in
Fig.~\ref{Histo071}.  In all cases, one can see that the distribution
of $\Delta_{AB}$ has acquired additional peaks compared to the
distribution for $\Delta$.  These arise because, during the quench
from $\phi_i$ to $\phi$, in some of the copies of the system one or
more disks belonging to defects have managed to escape their cages and
have crossed to the other side of the channel.  The splitting of the
peaks due to channel crossing would be approximately $h^2/N$, which is
consistent with the data in Fig.~\ref{Histo071}.  It should be
understood that this process is possible even though the compression
time is much shorter than $\tau_\alpha$: Fig.~\ref{Histo071} shows
that disks have crossed in only a few of the copies of the system,
despite the presence of ${\sim}10$ defects in each copy.  Our
explanation of why there is a splitting of $\Delta$ and $\Delta_{AB}$
does not invoke a Gardner transition and is similar to that given in
Ref.~\cite{scalliet:17} for soft spheres, except that in our system we
can explicitly identify the localized defects associated with the
disks that move.

Nevertheless, evidence for an avoided Gardner transition in our
hard-disk system can be found by studying the appropriate correlation
length.  Following Ref.~\cite{Berthier:16}, we define this from the
large-distance behavior of the correlation function associated with
$\Delta_{AB}^i=(y_{A,i}-y_{B,i})^2$, i.e., $G_P^0(k)=\langle u_i u_{i+k}
\rangle$, where $u_i=\Delta_{AB}^i/\Delta_{AB}-1$ and $\Delta_{AB}$ is
the value on the plateau which is given by Eq.~(\ref{Deltapredict}).
For our system of disks moving in a narrow channel, $G_P^0(k)$ can be
obtained from a transfer-matrix calculation in the high-density regime
where channel crossing can be ignored \cite{Suppl:TMMethods}.

\begin{figure}
  \begin{center}
    \includegraphics[width = 3.5in]{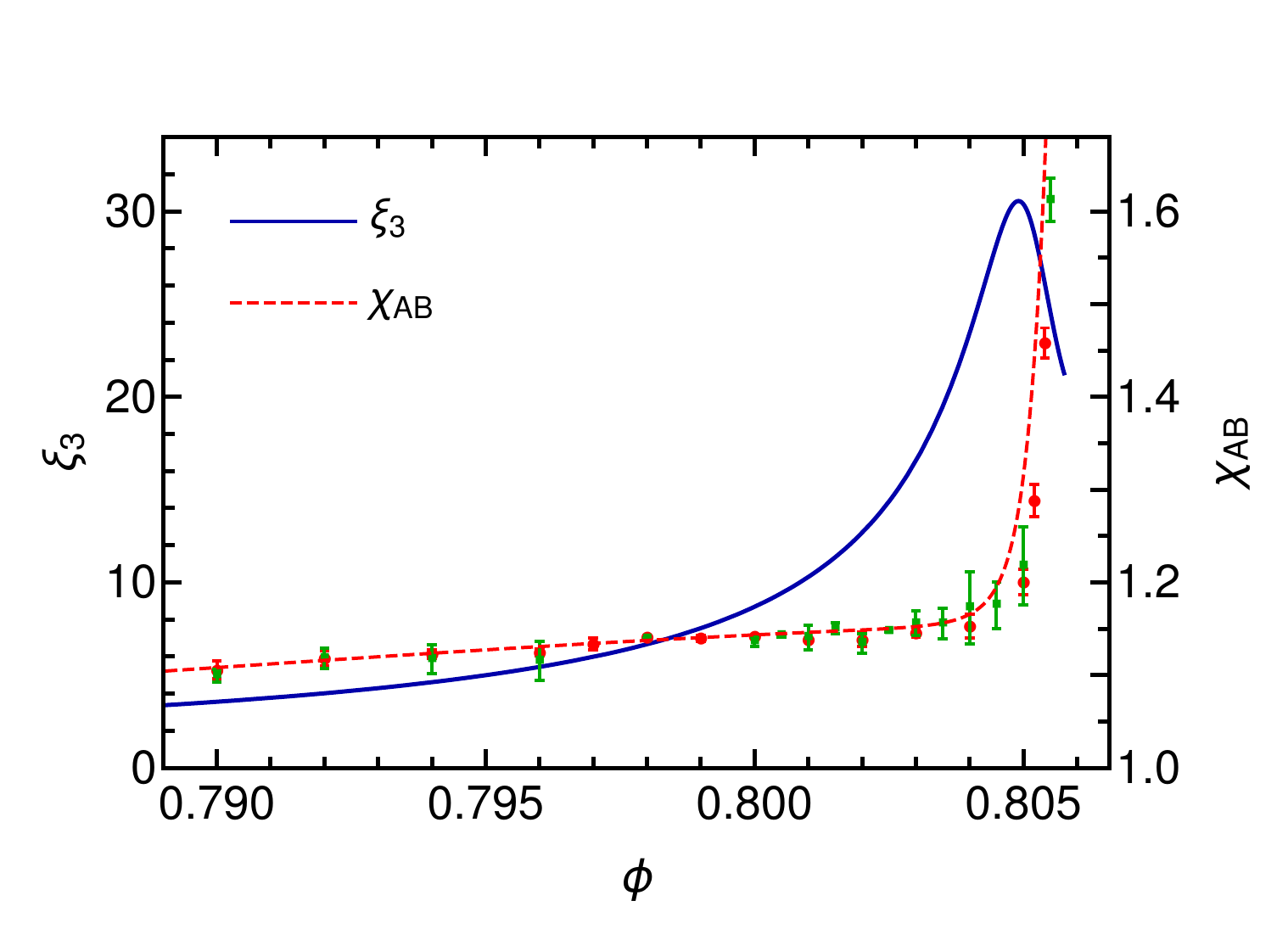}
    \caption{(Color online) The correlation length $\xi_3$ as a
      function of the packing fraction $\phi$, and the susceptibility
      $\chi_{AB}$, calculated using the transfer matrix procedure.
      Data points show values of $\chi_{AB}$ obtained in quenches from
      a state with no defects, starting from $\phi_i=0.76$ (red
      circles) and $\phi_i=0.78$ (green squares).}
    \label{fig:corrlens}
  \end{center}
\end{figure}

The correlation function shows a complicated behavior for small
separations $k$, but for large $k$ it decreases exponentially as
$G_P^0(k) \sim (-1)^k \exp[-k/\xi_3]$.  In Fig.~\ref{fig:corrlens} we
show how the correlation length $\xi_3$ depends on the packing
fraction $\phi$.  The length scale $\xi_3$ (which can also be
calculated from the eigenvalues of the transfer matrix
\cite{Godfrey:15,Suppl:TMMethods}) rises to its maximum value at
$\phi=0.8049$ before falling again.  Such behavior is typical of an
avoided transition.  Expressed as a distance, $\xi_3 L/N\approx
0.50\,\xi_3\sigma$, the maximum value of the correlation length is
approximately $15\sigma$.

The susceptibility $\chi_{AB}$ is defined, following
Ref.~\cite{scalliet:17}, as $\chi_{AB}=N
\operatorname{var}(\Delta_{AB})/ \operatorname{var}(\Delta_{AB}^i)$,
which is equivalent to
\begin{equation}
  \chi_{AB}=\frac{\Delta_{AB}^2}{\operatorname{var}(\Delta_{AB}^i)}
  \sum_{k=-\infty}^\infty G_P^0(k).
\label{gp:def}
\end{equation}
Figure~\ref{fig:corrlens} shows that $\chi_{AB}$ grows rapidly as the
density increases, with no sign of levelling off as might have been
expected for an avoided transition.  This is because the sum in
Eq.~(\ref{gp:def}) is dominated by the small $k$ region (see the
Supplemental Material \cite{Suppl:TMMethods}).

To summarize, the splitting between the plateau values of $\Delta$ and
$\Delta_{AB}$ does not provide strong evidence for a Gardner
transition, as it is an artifact that arises when there is a
significant chance that a disk will escape its cage during the quench.
A similar explanation may apply to the observations reported in
Refs.~\cite{Berthier:16,Seguin:16}.  Our study of the correlation
length $\xi_3$ shows that the Gardner transition is an avoided
transition for our one-dimensional system of hard disks.  The large
magnitude of the correlation length at the avoided transition suggests
that for hard spheres in three dimensions the equivalent length scale
could be very large; thus we anticipate that it will be challenging
for simulations to distinguish an avoided Gardner transition from a
true phase transition.  Nevertheless, we expect the Gardner transition
to be an avoided transition for any potential, hard or soft, in
dimensions $d \le 6$.

\begin{acknowledgments}
We should like to thank Ludovic Berthier and Francesco Zamponi for
sharing their insights.
\end{acknowledgments}



\clearpage
\begin{center}
	\textbf{\large Supplemental Material}
\end{center}







This document provides information on how the transfer-matrix method
of equilibrium statistical physics \cite{Baxter:82} can be used to
calculate some non-equilibrium properties of a system of confined hard
disks.  In Sec.~\ref{sec:timescale} we explain how the timescale for
$\alpha$ relaxation can be estimated and in
Secs.~\ref{sec:correlation} and \ref{sec:susceptibility} we show that
the correlation function $G_P^0(k)$ and the related susceptibility
$\chi_{AB}$ can be approximated by making use of correlation functions
calculated for a single copy of the system.

\section{Timescale for \texorpdfstring{$\alpha$}{alpha} relaxation}
\label{sec:timescale}

\begin{figure}[ht]
  \begin{center}
    \includegraphics[width = \columnwidth]{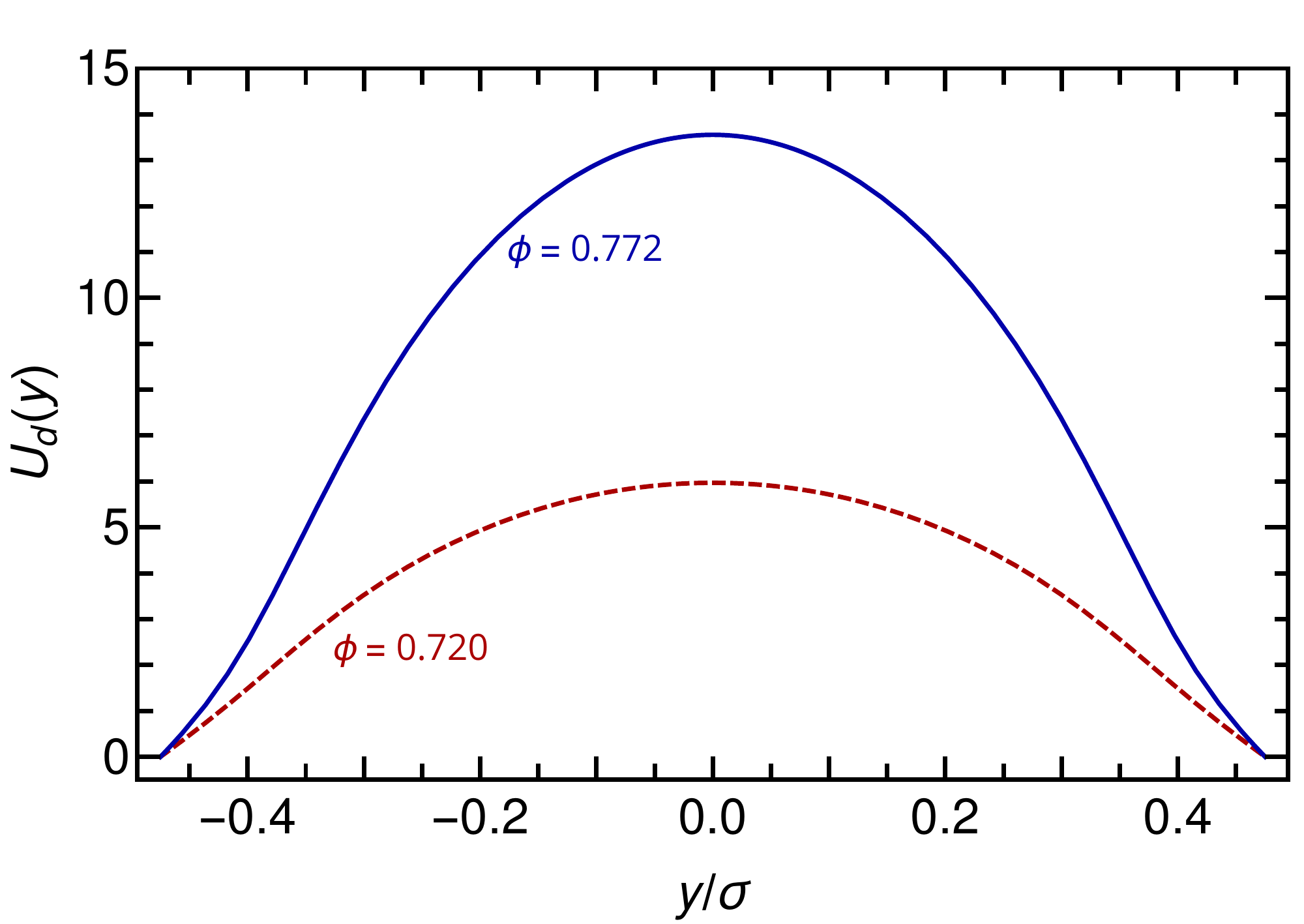}
    \caption{Free energy $U_d(y)$ of a disk that forms part of a
      defect as a function of its coordinate $y$.  Results, calculated
      from Eqs.~(\ref{Udofy}) and (\ref{rhoofy}) with $\beta=1$ and
      $h=0.95\sigma$, are shown for two values of the packing
      fraction, $\phi=0.720$ and $0.772$.  The height of the barrier
      increases with $\phi$, which leads to the large increase in the
      timescale $\tau_\alpha$ reported in the text.}
    \label{barrier}
  \end{center}
\end{figure}

For our system, the $\alpha$ relaxation time $\tau_{\alpha}$ is
determined by the motion of defects; that is, the length of the
plateau in $\Delta(t, t_W)$ is the typical time a disk that forms part
of a defect waits before crossing the channel.  We estimate the
timescale for this activated process from
\begin{equation}
  \tau_{\alpha}=\tau_0\, e^{\beta\Delta U_d},
  \label{taualpha}
\end{equation}
where $\beta$ is the reciprocal temperature, $\Delta U_d$ is the
height of the free energy barrier, $\tau_0$ is a microscopic timescale
of order $d_w/v_{\rm rms}$, $d_w$ is the average distance of a disk
from the wall, and $v_{\rm rms}$ is the root-mean-squared speed of the
disks.  We obtain an effective potential $U_d(y)$ from the
distribution function for the coordinate $y_2$ of a disk that is part
of a defect, so that its neighbors at $y_1$ and $y_3$ lie near
opposite sides of the channel; that is,
\begin{equation}
  \rho_d(y_2)\propto e^{-\beta U_d(y_2)},
  \label{Udofy}
\end{equation}
where the distribution of $y_2$ is given by
\begin{equation}
  \rho_d(y_2)= A\!\int_0^{h/2}\!\!\!\int_{-h/2}^0
  \rho(y_1,y_2,y_3)\,dy_1\, dy_3\,.
  \label{rhoofy}
\end{equation}
In Eq.~(\ref{rhoofy}), $\rho(y_1,y_2,y_3)$ is the $3$-disk
distribution function, calculated using the transfer matrix method of
Ref.~\cite{Godfrey:15}, and $A$ is a normalization constant,
independent of $y_2$.  Figure~\ref{barrier} illustrates the form of
$U_d(y_2)$ found in this way.  We estimate the barrier height $\Delta
U_d$ from $\Delta U_d= U_d(0)-U_d(h/2-d_w)$.  For $\phi = 0.720$ and
$0.745$, we find $\tau_{\alpha}\approx 5$ and $35$, respectively:
these values are consistent with the molecular-dynamics results shown
in Fig.~2.  For $\phi=0.772$ we obtain $\tau_\alpha\approx6900$, which
is greater than observation time in Fig.~2.

\section{Spatial correlations}
\label{sec:correlation}

The correlation function $G_P^0(k)$ is defined in terms of the
relative fluctuations of
$\Delta_{AB}^i\equiv(y_{A,i}-y_{B,i})^2$ away from their
average value on the plateau, which in this Section we denote simply
by $\Delta_{AB}$; i.e.,
\begin{equation}
  G_P^0(k) = \langle u_iu_{i+k}\rangle,
  \label{GPz}
\end{equation}
where $u_i = (\Delta_{AB}^i-\Delta_{AB})/\Delta_{AB}$.  In
Eq.~(\ref{GPz}), the average is over many different initial positions
for the disks, such that $y_{A,i}(t=0)=y_{B,i}(t=0)$.

The long correlation length for zigzag order and the long relaxation
times at high packing fractions allow the two-copy correlation
function to be approximated by using results from a single-copy
transfer-matrix calculation.  The method we use is a straightforward
extension of that described in the main text, where it is argued that
the plateau value of $\Delta_{AB}(t)$ can be approximated by
$2(\expect{y^2}-\expect{|y|}^2)$, the average being taken with respect
to the equilibrium state of a single copy.

For times $t$ that are small compared with the channel-crossing time
$\tau_\alpha$, the signs of $y_{A,i}(t)$ and $y_{B,j}(t)$ are
determined by the pattern of defects in the zigzag order, which is the
same in each copy.  We suppose that if the time is large compared with
the $\beta$ relaxation time $\tau_\beta$, $y_{A,i}(t)$ and
$y_{B,j}(t)$ will be statistically independent, apart from their
definite sign relation; naturally, this requires the relevant
timescales to be well separated, such that $\tau_\beta\ll
t\ll\tau_\alpha$.  Thus, a term such as
$\expect{y_{A,i}^2y^{}_{A,i+k}y^{}_{B,i+k}}$, which arises in the
expansion of $G_P(k)$, can be replaced by
$\expect{y_{A,i}^2|y^{}_{A,i+k}|}\expect{|y^{}_{B,i+k}|}\equiv\expect{y_i^2|y_{i+k}|}\expect{|y_{i+k}|}$,
in which the copy labels $A$ and $B$ can be omitted, as the averages
are the same for each copy.  The final simplification is to
approximate each single-copy average by an average over the
translationally-invariant equilibrium state, so that
$\expect{y_i^2|y_{i+k}|}\expect{|y_{i+k}|}\approx\expect{y_0^2|y_k|}\expect{|y|}$.
We expect this approximation to work well at high density, where the
spacing between defects is very large.

\begin{figure}[ht]
  \begin{center}
    \includegraphics[width = \columnwidth]{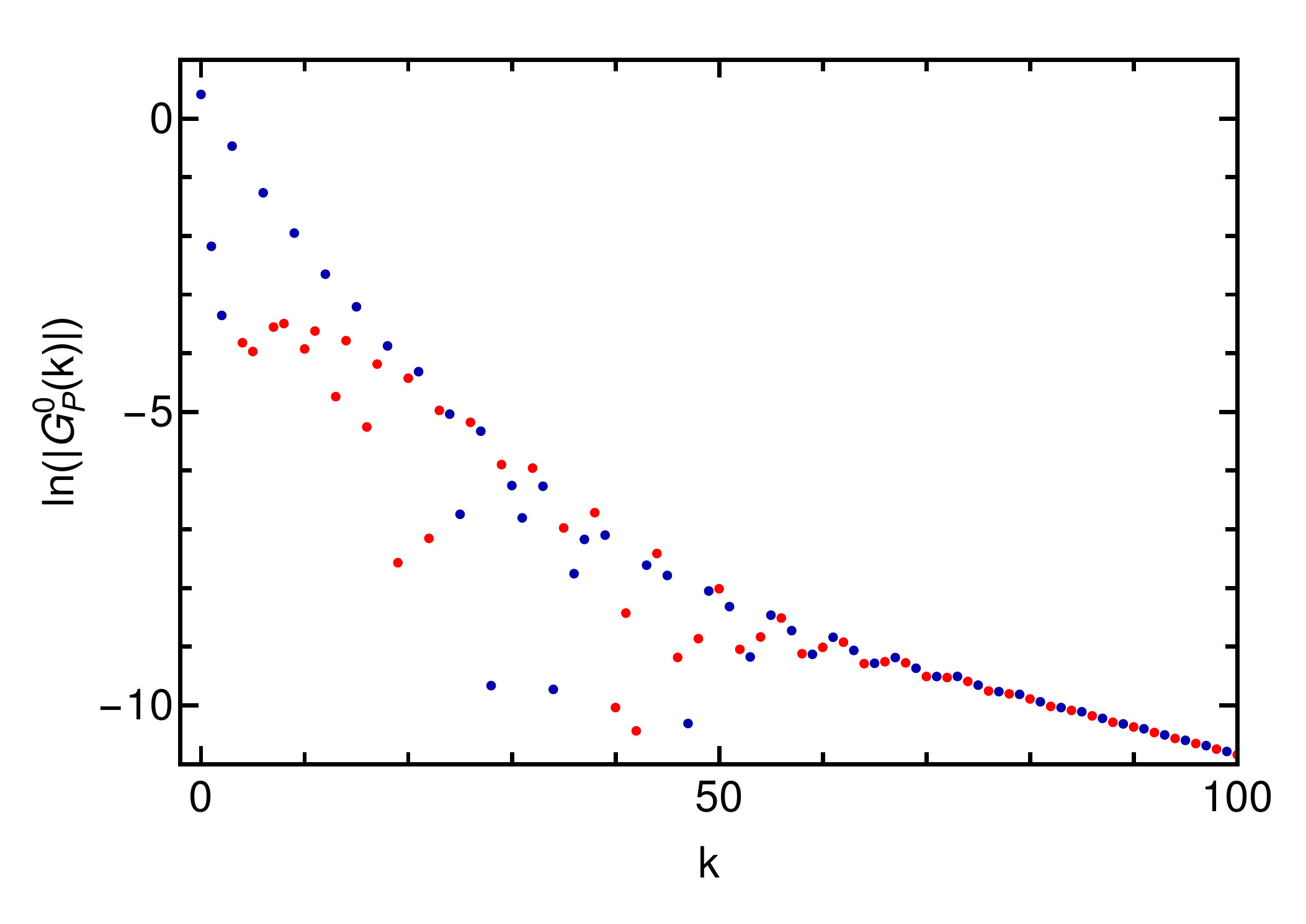}
    \caption{The correlation function $G_P^0(k)$ as a function of disk
      separation $k$ at a packing fraction $\phi= 0.8058$.  Blue dots
      indicate that $G_P^0(k)$ is positive, red dots that it is
      negative.  The non-monotonic decrease of $\lvert G_P^0(k)\rvert$
      for small $k$ is due to short-ranged crystalline correlations.
      For large $k$, the gradient of this log--linear plot is
      $-1/\xi_3$, which gives the correlation length $\xi_3\simeq21$.}
    \label{corrfun}
  \end{center}
\end{figure}

By applying the method outlined above, we find
\begin{align}
 G_P(k) \approx {}&\bigl\{2\expect{y_0^2y_{k}^2}+ 4\expect{|y_0||y_{k}|}^2  \nonumber\\
  & -8\expect{y_0^2|y^{}_{k}|}\expect{|y|}+2\expect{y^2}^2\bigr\}/\Delta_{AB}^2 - 1,
\end{align}
which depends on correlation functions of the form
$\expect{|y_0|^m|y_k|^n}$.  Equilibrium correlation functions are
calculated from the transfer matrix by the method discussed in Sec.~V
of Ref.~\cite{Godfrey:15} (described there for the case $k=1$) and in
Sec.~2.2 of Ref.~\cite{Baxter:82} (for the one-dimensional Ising
model).  Results from this calculation are shown in Fig.~\ref{corrfun}
for a case where the packing fraction is large, $\phi=0.8058$.

Figure~\ref{corrfun} show that the correlation function $G_P^0(k)$
decays rapidly as $k$ increases.  For $k<50$ this decay is
non-exponential and non-monotonic, owing to the presence of decaying,
period-3 oscillations; this can be seen most clearly for $k\le18$,
where a relatively large, positive value of $G_P^0$ is always followed
by two somewhat smaller values of either sign.  Short-ranged
crystalline correlations of this kind have been discussed in
Ref.~\cite{Godfrey:15}.

For sufficiently large separations, $k>70$, the decay of $G_P^0$ is
approximated well by an exponential, $G_P^0(k) \sim (-1)^k
\exp[-k/\xi_3]$, which allows us to extract the correlation
length~$\xi_3$.  (A similar method has been used in the Supplement to
Ref.~\cite{Berthier:16}, where a closely-related correlation length
$\xi_P$ was calculated for a system of hard spheres.)  For our system
of disks in a channel, we find that $\xi_3$ is the same correlation
length that determines the decay of correlations between the
$x$-separations of nearest-neighbor pairs of disks,
\begin{equation}
  \langle(x_{i+1}-x_i)(x_{i+k+1}-x_{i+k})\rangle_c \sim (-1)^ke^{-k/\xi_3},
  \label{eq:altsign}
\end{equation}
as described in Ref.~\cite{Godfrey:15}.  This asymptotic form is, in
fact, generic for the correlations between even functions of the
variables $y_i$, because the eigenfunction corresponding to the
third-largest eigenvalue, $\lambda_3$, has the same (even) parity as
the eigenfunction with the largest eigenvalue, $\lambda_1$.  The
right-hand side of \eqref{eq:altsign} is identical to
$(\lambda_3/\lambda_1)^k$, so that the lengthscale $\xi_3$ can be
calculated directly from the largest and third-largest eigenvalues of
the transfer matrix \cite{Godfrey:15}; the alternating sign of the
correlation function for large $k$ follows from the fact that
$\lambda_3$ is negative.  As $\phi$ increases, $\xi_3$ rises to its
maximum value at a packing fraction $0.8049$ before falling again, as
shown in Fig.~[7] of the main text.  Thus, if the growth of the
correlation length $\xi_3$ indicates the approach to the Gardner
transition, this transition is avoided, as one should expect for a
one-dimensional system.

\section{Susceptibility \texorpdfstring{$\chi_{AB}$}{}}
\label{sec:susceptibility}

In Ref.~\cite{scalliet:17}, Scalliet {\it et al.\null} have defined a
susceptibility $\chi_{AB}=N \operatorname{var}(\Delta_{AB})/
\operatorname{var}(\Delta_{AB}^i)$, in which all quantities are
evaluated on the plateau in $\Delta_{AB}(t)$.  It is straightforward
to verify that this definition is equivalent, in our problem, to
\begin{equation}
  \chi_{AB}=\frac{\Delta_{AB}^2}{\operatorname{var}(\Delta_{AB}^i)}
  \sum_{k=-\infty}^\infty G_P^0(k).
\label{gp:def2}
\end{equation}
The prefactor ensures that $\chi_{AB}\rightarrow1$ in the absence of
spatial correlations~\cite{scalliet:17}.  The plateau value of the
denominator $\operatorname{var}(\Delta_{AB}^i)$ can be approximated by
the same method that we used to find the plateau value of
$\Delta_{AB}$; we find
\begin{align}
  \operatorname{var}(\Delta_{AB}^i)
  \approx 2\expect{y^4}
  &+ 2\expect{y^2}^2 
  - 8\expect{|y|^3}\expect{|y|} \nonumber\\
  &+ 8\expect{y^2}\expect{|y|}^2
  - 4\expect{|y|}^4.
\end{align}
As can be seen from Fig.~7 of the main text, the results of molecular
dynamics simulations agree well with the calculated susceptibility,
and provide support for the approximations made for the correlation
functions in Sec.~\ref{sec:correlation}.  We note from Fig.~7 that
$\chi_{AB}$ grows rapidly as the packing fraction increases through
$\phi=0.8049$, the position of the maximum in the correlation
length~$\xi_3$.  Our definition of the correlation length is
unambiguous, but it will be noticed from Fig.~\ref{corrfun} that
$G_P^0(k)$ is very small in the region of $k$ where exponential decay
sets in.  The terms that contribute most to the sum in
Eq.~(\ref{gp:def2}) are, in fact, those with small values of $k$.  The
behavior of $G_P^0(k)$ for small $k$ is due to several eigenfunctions
whose eigenvalues have magnitudes that are less than, but still
comparable with, $|\lambda_3|$; these include the complex eigenvalues
that are responsible for the period-3 oscillations noted earlier, in
Sec.~\ref{sec:correlation}.

\bibliography{refs}

\end{document}